\newif\ifConference

\ifConference
\documentclass{article}
\usepackage{spconf,amsmath,epsfig}
\else
\documentclass[onecolumn,11pt,draftclsnofoot]{IEEEtran}
\fi
\usepackage{amsmath}
\usepackage{amsfonts}
\usepackage{amssymb}
\usepackage{graphicx}
\usepackage{cases}
\usepackage{subfigure}
\usepackage{cite}
\usepackage{comment}
\ifConference
\excludecomment{TechnicalReport}\includecomment{Conference} \else
\includecomment{TechnicalReport} \excludecomment{Conference}
\fi
\usepackage[margin=9pt,font=footnotesize,labelfont=bf,labelsep=colon,figurename=Fig.,format=plain]{caption}[2007/12/23]

\newcommand{\x}{{\bf x}}

\newcommand{\y}{{\bf y}}
\renewcommand{\v}{{\bf v}}
\newcommand{\A}{{\bf A}}

\newcommand{\V}{{\bf V}}
\newcommand{\Q}{{\bf Q}}

\newcommand{\U}{{\bf U}}
\newcommand{\D}{{\bf D}}

\renewcommand{\S}{{\bf S}}
\newcommand{\F}{{\bf F}}
\renewcommand{\D}{{\bf D}}

\newcommand{\mM}{{\mathcal{M}}}

\newcommand{\mF}{{\mathcal{F}}}

\newcommand{\bbZ}{{\mathbb{Z}}}

\newtheorem{theorem}{Theorem}

\newcommand{\expp}[1]{e^{#1}}

\newcommand{\supp}{\operatorname{supp} }
\newcommand{\sinc}{\operatorname{sinc} }

\def\yl{\y(\lambda)}
\def\yL{\y(\Lambda)}
\def\xl{\x(\lambda)}
\def\tx{{\tilde{x}}}
\def\xL{\x(\Lambda)}
\def\linL{{\lambda\in\Lambda}}

\def\mFz{\mF_0}

\newcommand{\beq}{\begin{equation}}
\newcommand{\eeq}{\end{equation}}

\newcommand{\fnyq}{{f_\textrm{NYQ}}}
\newcommand{\Tnyq}{{T_\textrm{NYQ}}}
\def\QEDclosed{\mbox{\rule[0pt]{1.3ex}{1.3ex}}}

\ifConference
\title{Efficient Sampling of Sparse Wideband Analog Signals}
\twoauthors{Moshe~Mishali and Yonina~C.~Eldar}{\thanks{This work was
supported in part by the Israel Science Foundation under Grant no.
1081/07 and by the European Commission in the framework of the FP7
Network of Excellence in Wireless COMmunications NEWCOM++
(contract no. 216715).}Department of Electrical Engineering\\
Technion - Israel Institute of Technology\\
Haifa, Israel 32000\\
\{moshiko@tx,yonina@ee\}.technion.ac.il}
{Joel~A.~Tropp}{\thanks{JAT was supported in part by DARPA and ONR.}
Applied \& Computational Mathematics\\
California Institute of Technology\\Pasadena, CA 91125-5000\\
jtropp@acm.caltech.edu}

\else
\title{
Efficient Sampling of Sparse Wideband Analog Signals
\thanks{This work was
supported in part by the Israel Science Foundation under Grant no.
1081/07 and by the European Commission in the framework of the FP7
Network of Excellence in Wireless COMmunications NEWCOM++
(contract no. 216715). JAT was supported in part by DARPA and ONR. This work has been submitted to the IEEE for
possible publication. Copyright may be transferred without notice,
after which this version may no longer be accessible. \newline
Moshe Mishali and Yonina C. Eldar are with the Technion---Israel
Institute of Technology, Haifa Israel. Emails:
moshiko@tx.technion.ac.il, yonina@ee.technion.ac.il. Joel A. Tropp
is with the Applied \& Computational Mathematics department of
California Institute of Technology, Pasadena, CA 91125-5000.
Email: jtropp@acm.caltech.edu}}
\author{Moshe Mishali, Yonina C.~Eldar and Joel A. Tropp}
\fi

\begin{document}

\maketitle

\begin{abstract}
Periodic nonuniform sampling is a known method to sample
spectrally sparse signals below the Nyquist rate. This strategy
relies on the implicit assumption that the individual samplers are
exposed to the entire frequency range. This assumption becomes
impractical for wideband sparse signals. The current paper
proposes an alternative sampling stage that does not require a
full-band front end. Instead, signals are captured with an analog
front end that consists of a bank of multipliers and lowpass
filters whose cutoff is much lower than the Nyquist rate. The
problem of recovering the original signal from the low-rate
samples can be studied within the framework of compressive
sampling. An appropriate parameter selection ensures that the
samples uniquely determine the analog input. Moreover, the analog
input can be stably reconstructed with digital algorithms.
Numerical experiments support the theoretical analysis.
\end{abstract}
\section{Introduction}
\label{sec:intro}

Radio frequency (RF) technology enables the modulation of a
narrowband signal by a high carrier frequency.  As a consequence,
manmade radio signals are often \emph{sparse}.  That is, they
consist of relatively small number of narrowband transmissions
spread across a wide territory of spectrum.  A convenient
description for these signals is the \emph{multiband model} where
the frequency support of a signal resides within several
continuous intervals in a wide spectrum but vanishes elsewhere.

It has become prohibitive to sample modern multiband signals
because their Nyquist rates may exceed the specifications of the
best analog-to-digital converters (ADCs) by orders of magnitude.
As a result, any attempt to acquire a multiband signal must
exploit its structure in an intelligent way.


Previous work on multiband signals has shown that it is possible
to reduce the sampling rate by acquiring samples from a periodic
but nonuniform grid \cite{Vaidyanathan}.  Multi-coset sampling, a
specific strategy of this type, was analyzed in \cite{Bresler00},
which established that exact recovery is possible when the band
locations are known.  The blind case, in which the band locations
are unknown, has been extensively studied in \cite{MishaliSBR}.
Unfortunately, the sampling front ends proposed in
\cite{Vaidyanathan,Bresler00,MishaliSBR} are impractical for
wideband applications because they require ADCs whose sampling
rate is matched to the Nyquist rate of the input signal, even when
the average sampling rate is much lower.  Other limitations are
described in Section~\ref{sec:multicoset}.  Another recent
work~\cite{horowitz} has partially overcome these shortcomings
using a hybrid optic--electronic system at the expense of size and
cost.


In this paper, we analyze a practical sampling system inspired by
the recent work on the random demodulator~\cite{laska07}. This
system multiplies the input signal by a random square wave
alternating at the Nyquist rate, then it performs lowpass
filtering, and samples the signal at a lower rate. Our system
consists of a bank of random demodulators running in parallel.  We
show that, for an appropriate choice of parameters, our system
uniquely and stably determines a multiband input signal. Moreover,
we describe digital algorithms for reconstructing the signals from
the parallel samples.


We continue with an outline of the paper.
Section~\ref{sec:background} reviews essential background
material. In Section~\ref{sec:mixing}, we describe the system
design and a frequency-domain analysis that leads to an infinite
measurement vectors (IMV) system. Applying ideas from
\cite{MishaliReMBo}, we reduce the problem of locating the
frequency bands to a finite-dimensional compressive sampling
problem.  We then derive an appropriate choice of parameters for
the sampling system. Section~\ref{sec:sim} presents our numerical
experiments, which demonstrate that the system permits stable
signal recovery in the presence of noise.


\section{Formulation and Background}
\label{sec:background}

\subsection{Design Goals for Efficient Sampling}

\begin{figure}
\centering
\includegraphics[scale=0.65]{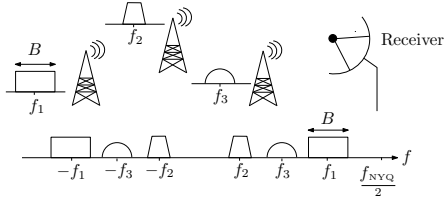}
\caption{Three RF transmissions with different
carriers $f_i$.  The receiver sees a multiband signal
(bottom drawing).  In this example $N=3$, and the
modulation techniques of the transmitters determine the maximal
expected width $B$.}\label{fig:TypicalM}
\end{figure}

Let $x(t)$ be a real-valued, finite-energy, continuous-time
signal, and let $X(f) = \int_{-\infty}^{\infty}x(t)\exp(-j2\pi
ft){\rm d}t$ be its Fourier transform.  We treat a multiband
signal model $\mM$ in which $x(t)$ is bandlimited to
$\mF=[-\fnyq/2,\fnyq/2]$ and the support of $X(f)$ consists of at
most $2N$ frequency interval whose widths do not exceed $B$.
Fig.~\ref{fig:TypicalM} depicts a typical communication
application that obeys this signal model.


We wish to design a sampling system for signals from our~model
$\mM$ that satisfies the following properties:
\begin{enumerate}
\item The sampling rate should be as low as possible;
\item the system has no prior knowledge of band locations; and
\item the system can be implemented with existing devices.
\end{enumerate}
We will call this type of sampling stage \emph{efficient}.



The set $\mM$ is a union of subspaces corresponding to all
possible signal supports. Every $x(t)\in\mM$ lies in one of these
subspaces. Detecting the exact subspace, prior to sampling, may be
impossible or too expensive to implement. An efficient system
should therefore be blind, in the sense that band locations are
not assumed to be known.

The lowest (average) sampling rate that allows blind perfect
reconstruction for all signals in $\mM$ is $4NB$ samples/sec
\cite{MishaliSBR}.  This rate is proportional to the effective
bandwidth of $x(t)$, and it is typically far less than the Nyquist
rate $\fnyq$, which depends only on the maximum frequency in
$x(t)$. See Section~\ref{sec:stability} for more discussion.


Our previous work describes blind reconstruction of $x(t)\in\mM$
from multi-coset samples taken at the minimal rate
\cite{MishaliSBR}. The next section details the practical
limitations of the multi-coset strategy, which make it inefficient
for wideband signals.


\subsection{Practical Limitations of Multi-Coset
Sampling}\label{sec:multicoset}

Multi-coset sampling involves periodic nonuniform sampling of the
Nyquist-rate sequence $x(n\Tnyq)$, where $\Tnyq=1/\fnyq$. The
$i$th coset takes the $i$th value in every block of $L$
consecutive samples. Retaining only $p<L$ cosets, indexed by
$C=\{c_i\}_{i=1}^p$, gives $p$ sequences
\begin{numcases}{x_{c_i}[n]=}
\nonumber x(n\Tnyq) & $n=mL+c_i, m\in\mathbb{Z}$  \\
 0 & otherwise, \label{xci}
\end{numcases}
with an average sampling rate $p/({L\Tnyq})$, which is lower than
the Nyquist rate.

To explain the practical limitations of this strategy, we observe
that standard ADC devices have a specified maximal rate $r$, and
manufactures require a preceding low-pass filter with cutoff
$r/2$. Distortions occur if the anti-aliasing filter is not used,
since the design is tailored to $r/2$-bandlimited signals and has
an internal parasitic response to frequencies above $r/2$. To
avoid these distortions, an ADC with $r$ matching the Nyquist rate
of the input signal must be used, even if the actual sampling rate
is below the maximal conversion rate $r$. In multi-coset sampling,
each sequence $x_{c_i}[n]$ corresponds to uniform sampling at rate
$1/(L\Tnyq)$, whereas the input $x(t)$ contains frequencies up to
$\fnyq/2$. Acquiring $x_{c_i}[n]$ is only possible using an ADC
with $r=\fnyq$, which runs $L$ times slower than its maximal rate.
Besides the resource waste, this renders multi-coset sampling
impractical in wideband applications where $\fnyq$ is higher
(typically by orders of magnitude) than the rate $r$ of available
devices.

One recent paper \cite{horowitz} developed a nonconventional ADC
design for wideband applications by means of high-rate optical
devices. The hybrid optic--electronic system allows sampling at
rate $1/(L\Tnyq)$ with minimal attenuation to higher frequencies
(up to $\fnyq/2$). Unfortunately, at present, this performance
cannot be achieved with purely electronic technology. Thus, for
wideband applications that cannot afford the size or expense of an
optical system, multi-coset sampling becomes impractical.

Another limitation of multi-coset sampling, which also exists in
the optical implementation, is maintaining accurate time delays
between the ADCs of different cosets. Any uncertainty in these
delays hobbles the recovery from the sampled sequences.

Before describing the way our proposed sampling stage overcomes
these limitations, we briefly review the mechanism underlying the
blind reconstruction of \cite{MishaliSBR}.

\subsection{IMV System}

Let $\xL = \{\xl : \linL\}$ be a collection of $n$-dimensional
vectors indexed by a fixed set $\Lambda$ that may be infinite. The
\emph{support} of a vector is the set $\supp(\v)=\{i\,|\,\v_i\neq
0\}$, and we define $\supp(\xL)=\cup_\lambda \supp(\xl)$.  We will
assume that the vectors in $\xL$ are jointly $K$-sparse in the
sense that $|\supp(\xL)| \leq K$.  In words, the nonzero entries
of each vector $\xl$ lie within a set of at most $K$ indices.


\begin{figure}
\centering
\includegraphics[scale=0.65]{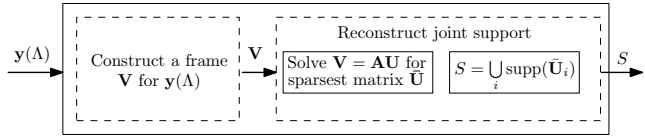}
\caption{Recovery of the joint support
$S=\supp(\xL)$.}\label{fig:CTF}
\end{figure}

Let $\A$ be an $m\times n$ matrix with $m<n$, and consider a
parameterized family of linear systems
\begin{equation}\label{eq:IMV}
\yl=\A \xl,\quad\linL.
\end{equation}
When the support $S=\supp(\xL)$ is known, recovering $\xL$ from
the known vector set $\yL=\{\yl:\linL\}$ is possible if the
submatrix $\A_S$, consisting of the columns of $\A$ indicated by
$S$, has full column rank. In this case,
\begin{subequations}\label{eq:recons}
\begin{equation}
\x_S(\lambda) =(\A_S)^\dag \yl
\end{equation}
\begin{equation}
\x_i(\lambda)=0,\quad i\notin S
\end{equation}
\end{subequations}
where $\x_S(\lambda)$ contains only the entries of $\x$ indexed by
$S$, where $\A_S^H$ denotes the conjugate transpose of $\A_S$ and
where $(\A_S)^\dag = ( \A_S^H \A_S)^{-1} \A_S^H$ is the
Moore--Penrose pseudoinverse.  For unknown support $S$,
(\ref{eq:IMV}) is still invertible if $K=|S|$ is known, and every
set of $2K$ columns from $\A$ is linearly independent
\cite{MElad,Chen,MishaliReMBo}.  In general, solving
(\ref{eq:IMV}) for $\xL$ is NP-hard because it may require a
combinatorial search. Nevertheless, recent advances in compressive
sampling and sparse approximation delineate situations where
polynomial-time recovery algorithms correctly identify
$\supp(\xL)$ for finite $\Lambda$. This challenge is sometimes
referred to as a multiple measurement vectors (MMV) problem
\cite{CandesRobust,Donoho,Cotter,Chen,TroppI,TroppII}.

%

Recovering a multiband signal $x(t)$ from a set of multi-coset
samples can be reduced to a certain infinite measurement vectors
(IMV) problem (where $\Lambda$ is infinite).  When the band
locations are known, the support set $S$ is determined and
reconstruction can be performed via
(\ref{eq:recons})~\cite{Vaidyanathan,Bresler00}.  In a blind
scenario, the support of the unknown vectors $\xl$ can be
recovered in two steps~\cite{MishaliSBR,MishaliReMBo}.  First,
construct a (finite) frame $\V$ for $\yL$.  Then, find the
(unique) solution $\bar{\U}$ to the MMV system $\V = \A\U$ that
has the fewest nonzero rows.  It holds that $\supp(\xL)$ is the
set $S = \cup_i \supp(\bar{\U}_i)$, where the union occurs over
columns of $\bar{\U}$.  Fig.~\ref{fig:CTF} summarizes these
recovery steps.



In the next section, we describe and analyze the proposed sampling
system. In contrast to the multi-coset strategy, our system uses
standard low-rate ADCs. We match the analog input of the ADCs to
their maximal rate. The system also avoids time offsets between
devices.  As in multi-coset sampling, the sampling sequences
generated by our system are related to $x(t)$ via an IMV system,
different from the one which is based on the sequences
(\ref{xci}). Consequently, the recovery of $x(t)$ can be performed
via the steps described in Fig.~\ref{fig:CTF} and
(\ref{eq:recons}).


\section{Efficient Sampling}
\label{sec:mixing}

\subsection{Description}

Let us present the proposed system in more detail. A block diagram
appears in Fig.~\ref{fig:OurSystem}. We discuss the choice of
system parameters in the sequel.


\ifConference\else
\begin{figure*}
\centering \mbox {
\includegraphics[scale=0.70]{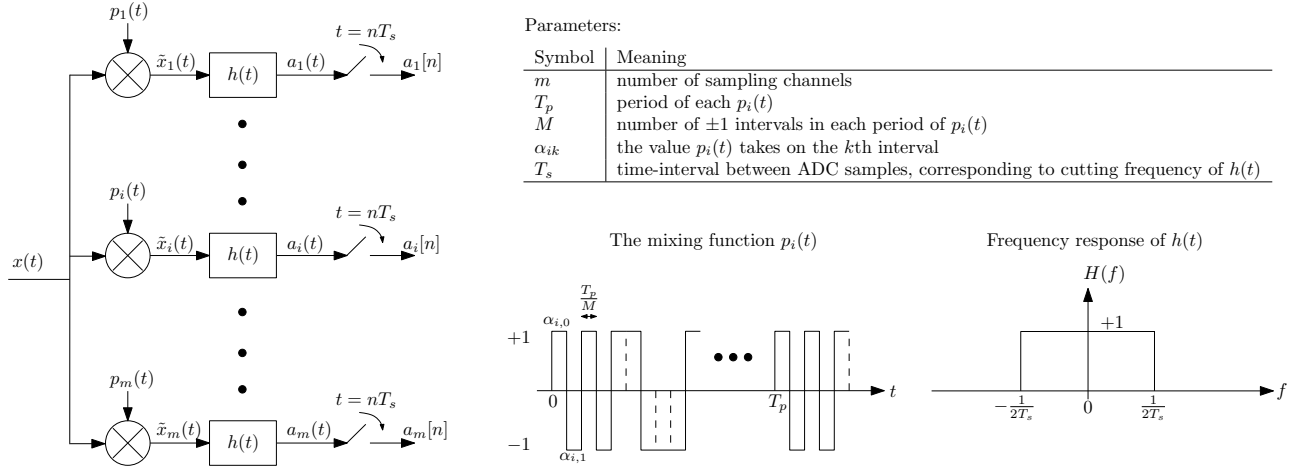}}
\caption{Description of a practical sampling stage for multiband
signals.} \label{fig:OurSystem}
\end{figure*}
\fi

The signal $x(t)$ enters $m$ channels simultaneously.  In the
$i$th channel, $x(t)$ is multiplied by a \emph{mixing function}
$p_i(t)$, which is a $T_p$-periodic, piecewise constant function
that alternates between the levels $\pm 1$ for each of $M$ equal
time intervals.  Formally,
\begin{equation}
p_i(t)=\alpha_{ik},\quad k\frac{T_p}{M} \leq t \leq
(k+1)\frac{T_p}{M},\quad 0\leq k\leq M-1,
\end{equation}
with $\alpha_{ik}\in\{+1,-1\}$, and $p_i(t+nT_p) = p_i(t)$ for
every $n\in\bbZ$.


After mixing, the output is converted to digital using the
standard approach.  In each channel, the signal spectrum is
truncated by a lowpass filter with cutoff $1/({2T_s)}$ and the
filtered signal is sampled at rate $1/T_s$.

Note that the cutoff and the sampling rate match, and each channel
operates independently.  Since there are $m$ channels, the average
sampling rate is $m/T_s$ samples/sec.  A further advantage of this
type of system is that samples are produced at a constant rate, so
they may be fed to a digital processor operating at the same
frequency, whereas multi-coset sampling requires an additional
hardware buffer to synchronize the nonuniform sequences.


\subsection{Analysis}

To ease exposition we choose an odd $M$, $T=M/\fnyq$, and
$T_s=T_p=T$. \ifConference
\else These choices are relaxed in the
sequel. \fi~Consider the $i$th channel. Since $p_i(t)$ is
periodic, it has a Fourier expansion \ifConference\else
\begin{align}
p_i(t)=\sum_{n=-\infty}^\infty c_{in} \expp{j\frac{2\pi}{T}nt},
\end{align}
where
\begin{align}
c_{in} & = \frac{1}{T}\int_0^{T}p_i(t)\expp{-j\frac{2\pi}{T}nt}dt \\
& = \frac{1}{T} \int_0^{T/M} \sum_{k=0}^{M-1} \alpha_{ik} \expp{-j\frac{2\pi}{T}n(t+k\frac{T}{M}}dt\\
& = \frac{1}{T} \sum_{k=0}^{M-1} \alpha_{ik}
\expp{-j\frac{2\pi}{M}nk}\int_0^{T/M}\expp{-j\frac{2\pi}{T}nt}dt.\label{eq:cn}
\end{align}
Evaluating the integral gives
\begin{align}
c_{in} =
\frac{1}{2\pi}\left(\sum_{k=0}^{M-1}\alpha_{ik}\expp{-j\omega_0
nk}\right)\frac{1-\expp{-j\omega_0 n}}{jn},
\end{align}
where $\omega_0 = 2\pi/M$ and $c_{in}=c_{i,-n}$.
\fi
~Expressing the
Fourier transform $P_i(f)$ in terms of the Fourier series
coefficients $c_{in}$ leads to
\begin{align}
P_i(f) &= \int_{-\infty}^{\infty} p_i(t)\expp{-j2\pi ft}dt  =
\sum_{n=-\infty}^{\infty} c_{in}\delta\left(f-\frac{n}{T}\right),
\end{align}
with $\delta(t)$ denoting the Dirac delta function. The analog
multiplication $\tx_i(t) = x(t)p_i(t)$ translates to convolution
in the frequency domain,
\begin{align}
\tilde{X}_i(f) &= X(f)\ast P_i(f) = \sum_{n=-\infty}^\infty
c_{in}X\left(f-\frac{n}{T}\right).
\end{align}
Therefore, $\tilde{X}_i(f)$ is a linear combination of shifted
copies of $X(f)$.

Filtering $\tilde{X}_i(f)$ by $H(f)$, whose frequency response is
an ideal rect function in the interval $\mFz = [-1/(2T),1/(2T)]$,
results in
\begin{align}\label{eq:AfterFilter}
A_i(f) & = H(f)\tilde{X}_i(f)  = \sum_{n=-n_0}^{n_0}
c_{in}X\left(f-\frac{n}{T}\right), \quad f\in\mFz,
\end{align}
where $n_0$ is the smallest integer satisfying \ifConference
\else
\begin{equation}
2n_0 + 1 \geq T\fnyq.\label{eq:n0}
\end{equation}\fi~
Under the choices above, $n_0 = (M-1)/2$. The discrete-time
Fourier transform of $a_i[n]$ is
\begin{align}
A_i(\expp{j 2 \pi f T}) &= \sum_{n=-\infty}^\infty a_i[n]\expp{-j2
\pi f T n}\label{eq:DTFT_ai} \\&= \sum_{n=-n_0}^{n_0} c_{in}
X\left(f-\frac{n}{T}\right),\quad f\in\mFz.\label{eq:OurRelation}
\end{align}
Substituting (\ref{eq:cn}) in (\ref{eq:OurRelation}) leads to the
system
\begin{align}\label{eq:OurIMV}
\y(f) &= (\S \F) (\D \x(f)),\quad f\in\mFz,
\end{align}
where $\y_i(f)=A_i(\expp{j 2 \pi f T}),\,1\leq i \leq m$, $\S$ is
an $m \times M$ matrix whose $ik$th entry $\S_{ik}=\alpha_{ik}$.
The $M \times M$ matrix $\F$ is a certain cyclic columns shift of
the discrete Fourier transform matrix of order $M$. The $M$-square
diagonal $\D$ scales
\begin{equation}\label{eq:xif}
\x_i(f)=X(f+(i-n_0-1)/T)
\end{equation}
according to the last term in (\ref{eq:cn}). Since $\D$ has
non-zero diagonal entries, it can be absorbed into $\x(f)$ while
keeping $\supp(\x(\mFz)) = \supp(\D\x(\mFz))$. Thus,
(\ref{eq:OurIMV}) is an IMV system with $\S\F$ replacing $\A$ of
(\ref{eq:IMV}).

\ifConference\else

We now explain the choices assumed in the beginning of the
section.
\begin{enumerate}
\item $T_s=T_p$ justifies the implication
    (\ref{eq:DTFT_ai})-(\ref{eq:OurRelation}). For $T_s<T_p$,
    (\ref{eq:OurRelation}) does not contain the entire
    information about $X(f)$. Thus, this choice should be
    avoided. In contrast, the selection $T_s>T_p$ has the
    following benefit for hardware implementation. Suppose
    $T_s=k T_p$ for some integer $k>1$. Then, it can be
    verified that under technical conditions on $\alpha_{ik}$,
    each sequence $a_i[n]$ corresponds to $k$ channels which
    are designed with $T_s=T_p$. Consequently, the number of
    mixers, filters and ADCs is reduced by a factor of $k$.
    This choice requires an ADC with $r=kT_p$ and is thus
    limited by the sampling rates of available devices.

\item Using $T=M/\fnyq$ ensures that the right hand side of
    (\ref{eq:n0}) is an integer. Other choices are possible,
    but imply a sampling rate that is higher than the minimum.

\item Odd $M$ simplify the exposition. However, a unique
    solution for the IMV (\ref{eq:OurIMV}) can also be
    guaranteed for even $M$. An even $M$ also requires slight
    modifications for the reconstruction algorithms of
    \cite{MishaliSBR} as will be detailed in the journal
    version of this paper.
\end{enumerate}

\fi

\subsection{Parameter Selection and Stable Recovery}
\label{sec:stability}

The following theorem suggests a parameter selection for which the
infinite sequences $a_i[n], 1\leq i \leq m$ match a unique
$x(t)\in\mM$. When the band locations are known, the same
selection works with half as many sampling channels.  Thus, the
system appearing in Fig.~\ref{fig:OurSystem} can also replace the
multi-coset stage of \cite{Bresler00}.


\begin{theorem}[Uniqueness]\label{th:uniq}
Let $x(t) \in \mM$ be a multiband signal and assume the choices
$T=M/\fnyq$ for an integer $M$ (not necessarily odd) and
$T_p=T_s=T$. If
\begin{enumerate}
\item $M \leq \fnyq/B$,
\item $m \geq 2N$ for non-blind reconstruction or $m\geq 4N$
    for blind,
\item $\S=\{\alpha_{ik}\}$ is such that every $4N$ columns are
    linearly independent,
\end{enumerate}
then, for every $f\in\mFz$, the vector $\x(f)$ is the unique
$2N$-sparse solution of (\ref{eq:OurIMV}). In addition, under
these choices $\x(\mFz)$ is jointly $4N$-sparse.
\end{theorem}

\begin{figure}
\centering \mbox {
\includegraphics[scale=0.45]{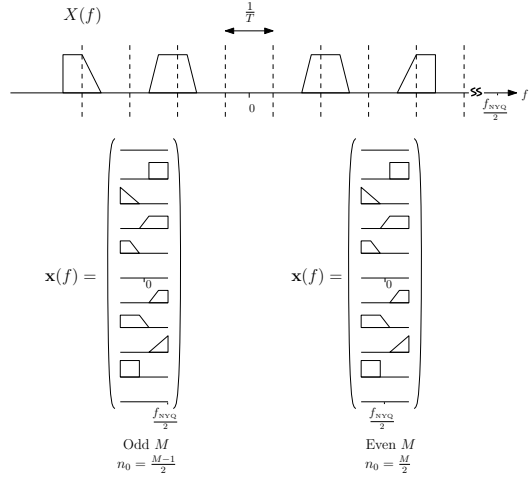}}
\caption{The relation between the Fourier transform $X(f)$ and the unknown vector set $\x(f)$, eq. (\ref{eq:xif}).} \label{fig:Xfxf}
\end{figure}

\textbf{Proof. } The proof goes along the line of
\cite{MishaliSBR}. The relation (\ref{eq:xif}) can be thought of
slicing the spectrum $X(f)$ into pieces of length $1/T$ and then
rearranging them in a vector form $\x(f)$. Fig.~\ref{fig:Xfxf}
visualizes this relation for even and odd $M$.

The choice $M \leq \fnyq/B$ ensures that $1/T\geq B$ and thus
every band can contribute only a single non-zero value to $\x(f)$.
As a consequence, $\x(f)$ is $2N$-sparse for every $f\in\mFz$. In
addition, this choice of $M$ and the continuity of the bands
guarantee that each band can occupy two spectrum pieces at the
most. Therefore, when aggregating the frequencies to compute
$S=\supp(\x(\mFz))$, we have $|S|\leq 4N.$

In the non-blind setting, the band locations imply the support set
$S$. The other two conditions on $m,\S$ ensure the existence of
$(\A_S)^\dag$, and thus (\ref{eq:recons}) provides the uniqueness
of $\x(f)$.

In blind recovery, $S$ is unknown, and the following CS result is
used to ensure the uniqueness. A $K$-sparse vector $\x$ is the
unique solution of $\y=\A\x$ if every $2K$ columns of $\A$ are
linearly independent \cite{MElad}. Clearly, this condition
translates to $m\geq 4N$ and the condition on $\S$ of the theorem.
{\hspace*{\fill}~\QEDclosed\par\endtrivlist\unskip}

The parameter selection of Theorem~\ref{th:uniq} guarantees an
average sampling rate $m/T \geq 4NB$. Depending on whether
$\fnyq/B$ is an integer, this selection allows to achieve the
minimal rate when taking the extreme values for $m,M$. Note that
each $\x(f)$ is $2N$-sparse, while $\x(\mFz)$ is jointly
$4N$-sparse under the parameter selection of the theorem. As
detailed in \cite{MishaliSBR}, this factor requires doubling $m$
in order to use Fig.~\ref{fig:CTF} and (\ref{eq:recons}). Gaining
back this factor at the expense of a higher recovery complexity is
also described in \cite{MishaliSBR}.

Verifying that a set of signs $\{ \alpha_{ik} \}$ satisfies the
requirement of the theorem is computationally difficult because
one must check the rank of every set of $4N$ columns from $\S$. It
is known that a random choice of signs will work, except with
probability exponentially small in $M$ \cite{TV06}.


In fact, recent work on compressive sampling shows that a random
choice of signs ensures that signal acquisition is stable
\cite{CandesRobust}. A matrix $\A$ is said to have the
\emph{restricted isometry property} (RIP) of order $K$, if there
exists $0\leq\delta_K<1$ such that
\begin{equation}\label{eq:rip}
(1-\delta_K) \|\x\|^2  \leq \|\A\x\|^2 \leq (1+\delta_K) \|\x\|^2
\end{equation}
for every $K$-sparse vector $\x$ \cite{CandesRobust}.  When
$\A=\S\F$ satisfies the RIP of order $4N$, then the matrices
$\A_S$ and $(\A_S)^\dag$ are well conditioned for every possible
frequency subset $S\subseteq \mFz$ with $|S|\leq 2N$.
\ifConference
\else~It was proved in \cite{C08} that basis
pursuit can recover the $K$-sparse solution $\x$ of $\y=\A\x$ for
an underdetermined $\A$, if $\A$ has $\delta_{2K}<\sqrt{2}-1$. The
mean squared error of the recovery in the presence of noise or
model mismatch was also shown to be bounded under the same
condition. Similar conditions were shown to hold for other
recovery algorithms. In particular, \cite{EM08} proved a similar
argument for a mixed $\ell_2/\ell_1$ program, for the MMV setting.
Thus, if $\A=\S\F$ has the RIP of order $4N$, then it implies the
stability of the recovery using Fig.~\ref{fig:CTF}, when the
mixed-norm program is utilized to solve the sparsest solution of
the MMV $\V=\A\U$.\fi


It remains to quantify when stable recovery is possible for
specific choices of $m$ and $M$.  Let $\F$ be an $M \times M$
unitary matrix (such as $\F$ in (\ref{eq:OurIMV})), and suppose
that $\S$ is an $m \times M$ random matrix whose entries are
equally likely to be $\pm 1/\sqrt{m}$. The RIP of order $K$ holds
with high probability for the matrix $\A = \S\F$ when $m \geq C K
\log(M/K)$, where $C$ is a positive constant independent of
everything \cite{SignsRIP}. The log factor is
necessary~\cite{BD08}. In practice, we empirically evaluate the
stability of the system since the RIP cannot be verified
computationally.


\section{Numerical Evaluation}
\label{sec:sim}

To evaluate the empirical performance of the proposed system (see
Fig.~\ref{fig:OurSystem}), we can simulate the action of the
system on test signals contaminated with white Gaussian noise. To
recover the signals from the sequences of samples, we apply the
reduction from an IMV system to an MMV system, as described in
Fig.~\ref{fig:CTF}.  We solve the resulting MMV systems using
simultaneous orthogonal matching pursuit~\cite{Cotter,TroppI}.

More precisely, we evaluate the performance on 100 noisy test
signals of the form $x(t) + w(t)$, where $x$ is a multiband signal
and $w$ is a white Gaussian noise process.  The multiband signals
consist of $N = 3$ pairs of bands, each of width $B = 40$ MHz,
constructed using the formula
$$
x(t)=\sum_{i=1}^N \sqrt{E_iB}\sinc(Bt)cos(2\pi f_i t),
$$
where $\sinc(x)=\sin(\pi x)/(\pi x)$.  The energy coefficients are
fixed $E_i=\{1,2,3\}$, whereas for every signal the carriers $f_i$
are chosen uniformly at random in $[-\fnyq/2,\fnyq/2]$ with $\fnyq
= 10 $ GHz.  To represent the continuous signals in simulation, we
place a dense grid of 4000 equispaced points in the time interval
$[-200/\fnyq, 200/\fnyq]$.  The Gaussian noise is added and scaled
so that the test signal has the desired signal-to-noise ratio
(SNR), where we define the SNR to be $10 \log(\|x\|/\|w\|)$.

We simulate the proposed system with $m = 51$ channels, where each
mixing function $p_i(t)$ alternated sign at most $M = 51$ times.
The sampling rate parameters are chosen so that $T_s = T_p =
M/\fnyq$. Each sign $\alpha_{ik}$ is chosen uniformly at random
and fixed for the duration of the experiment.  To simulate the
analog lowpass filter, we use a 50-tap digital FIR filter,
designed with the MATLAB command \texttt{h=fir1(50,1/M)}.  The
output of the filter is decimated to produce the sampled sequences
$a_i[n]$.

\begin{figure}
\centering \mbox {
\includegraphics[scale=0.50]{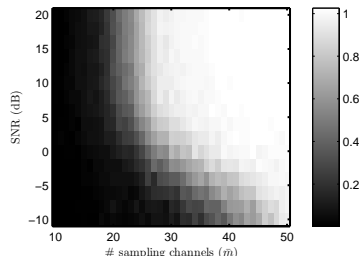}}
\caption{Image intensity represents percentage of correct support
set recovery $\hat{S}=S$, for reconstruction from different number
of sampling sequences $\bar{m}$ and under several SNR levels.}
\label{fig:sim_noise}
\end{figure}

The input signal is reconstructed from $\bar{m}\leq m$ channels.
We follow the procedure described in Fig.~\ref{fig:CTF} to obtain
an estimated support set $\hat{S}$. When $\hat{S} = S$, the true
support set, we declare that the system has recovered the signal.
Fig.~\ref{fig:sim_noise} reports the percentage of recoveries for
various numbers $\bar{m}$ of channels and various SNRs.


To construct the frame $\V$, we begin by computing the $m^2$
values $\Q_{ik} = \sum_n a_i[n] a_k[n]$.  We then perform the
eigenvalue decomposition $\Q = \V \V^H$ and then discard the
eigenvectors of the noise space \cite{MishaliSBR}.

\section{Conclusions}
\label{sec:conclusions}


We developed an efficient sampling stage for analog multiband
signals. In the proposed system, analog mixers and standard ADCs
replace impractical nonuniform sampling of multi-coset strategy.
Analog mixers for wideband applications is an existing RF
technology, though selecting the exact devices may require an
expertise in analog design.

The proposed system has a set of parameters, which determines the
signal, if selected according to the conditions we derived.
Analyzing our system in the frequency domain lead to an IMV
system, which allows to use existing reconstruction stages with
only minor modifications. In addition, based on the IMV system and
recent works in the CS literature, we deduce the rate requirements
for stable blind recovery, which in general is higher than the
rate required to determine the signal from its samples.

A preliminary computer evaluation of our system shows a promise
for stable blind recovery from sub-Nyquist sampling rate, although
further work is required to quantify the optimal working point in
the trade-off between sampling rate, blindness, and practical
implementation.

\bibliographystyle{IEEEtran}
\bibliography{IEEEabrv,Mishali_Eldar_Tropp_Eilat08}

\begin{thebibliography}{10}
\providecommand{\url}[1]{#1}
\csname url@rmstyle\endcsname
\providecommand{\newblock}{\relax}
\providecommand{\bibinfo}[2]{#2}
\providecommand\BIBentrySTDinterwordspacing{\spaceskip=0pt\relax}
\providecommand\BIBentryALTinterwordstretchfactor{4}
\providecommand\BIBentryALTinterwordspacing{\spaceskip=\fontdimen2\font plus
\BIBentryALTinterwordstretchfactor\fontdimen3\font minus
  \fontdimen4\font\relax}
\providecommand\BIBforeignlanguage[2]{{%
\expandafter\ifx\csname l@#1\endcsname\relax
\typeout{** WARNING: IEEEtran.bst: No hyphenation pattern has been}%
\typeout{** loaded for the language `#1'. Using the pattern for}%
\typeout{** the default language instead.}%
\else
\language=\csname l@#1\endcsname
\fi
#2}}

\bibitem{Vaidyanathan}
Y.-P. Lin and P.~P. Vaidyanathan, ``Periodically nonuniform sampling of
  bandpass signals,'' \emph{{IEEE} Trans. Circuits Syst. {II}}, vol.~45, no.~3,
  pp. 340--351, Mar. 1998.

\bibitem{Bresler00}
R.~Venkataramani and Y.~Bresler, ``Perfect reconstruction formulas and bounds
  on aliasing error in sub-nyquist nonuniform sampling of multiband signals,''
  \emph{{IEEE} Trans. Inform. Theory}, vol.~46, no.~6, pp. 2173--2183, Sep.
  2000.

\bibitem{MishaliSBR}
M.~Mishali and Y.~C. Eldar, ``Blind multi-band signal reconstruction:
  Compressed sensing for analog signals,'' \emph{CCIT Report no. 639, EE Dept.,
  Technion - Israel Institute of Technology; submitted to {IEEE} Trans. Signal
  Processing}, Sep. 2007.

\bibitem{horowitz}
M.~Fleyer, A.~Rosenthal, A.~Linden, and M.~Horowitz, ``Multirate synchronous
  sampling of sparse multiband signals,'' \emph{Arxiv preprint
  arXiv:0806.0579}, 2008.

\bibitem{laska07}
J.~N. Laska, S.~Kirolos, M.~F. Duarte, T.~S. Ragheb, R.~G. Baraniuk, and
  Y.~Massoud, ``Theory and implementation of an analog-to-information converter
  using random demodulation,'' in \emph{Proc. of. ISCAS 2007}, 2007, pp.
  1959--1962.

\bibitem{MishaliReMBo}
M.~Mishali and Y.~C. Eldar, ``Reduce and boost: {R}ecovering arbitrary sets of
  jointly sparse vectors,'' \emph{{IEEE} Trans. Signal Processing}, vol.~56,
  no.~10, pp. 4692--4702, Oct. 2008.

\bibitem{MElad}
D.~L. Donoho and M.~Elad, ``Maximal sparsity representation via $\ell 1$
  minimization,'' \emph{Proc. Natl. Acad. Sci.}, vol. 100, pp. 2197--–2202,
  Mar. 2003.

\bibitem{Chen}
J.~Chen and X.~Huo, ``Theoretical results on sparse representations of
  multiple-measurement vectors,'' \emph{{IEEE} Trans. Signal Processing},
  vol.~54, no.~12, pp. 4634--4643, Dec. 2006.

\bibitem{CandesRobust}
E.~J. Cand\`{e}s, J.~Romberg, and T.~Tao, ``{Robust uncertainty principles:
  Exact signal reconstruction from highly incomplete frequency information},''
  \emph{{IEEE} Trans. Inform. Theory}, vol.~52, no.~2, pp. 489--509, Feb. 2006.

\bibitem{Donoho}
D.~L. Donoho, ``Compressed sensing,'' \emph{{IEEE} Trans. Inform. Theory},
  vol.~52, no.~4, pp. 1289--1306, April 2006.

\bibitem{Cotter}
S.~F. Cotter, B.~D. Rao, K.~Engan, and K.~Kreutz-Delgado, ``Sparse solutions to
  linear inverse problems with multiple measurement vectors,'' \emph{{IEEE}
  Trans. Signal Processing}, vol.~53, no.~7, pp. 2477--2488, July 2005.

\bibitem{TroppI}
J.~A. Tropp, A.~C. Gilbert, and M.~J. Strauss, ``{Algorithms for simultaneous
  sparse approximation. Part I: Greedy pursuit},'' \emph{Signal Process.},
  vol.~86, pp. 572--–588, Apr. 2006.

\bibitem{TroppII}
J.~A. Tropp, ``{Algorithms for simultaneous sparse approximation. Part II:
  Convex relaxation},'' \emph{Signal Process.}, vol.~86, pp. 589--–602, Apr.
  2006.

\bibitem{TV06}
T.~Tao and V.~Vu, ``{On random $\pm$1 matrices: Singularity and determinant},''
  \emph{Random Structures Algorithms}, vol.~28, no.~1, pp. 1--23, 2006.

\bibitem{C08}
E.~Cand\`{e}s, ``The restricted isometry property and its implications for
  compressed sensing,'' \emph{C. R. Acad. Sci. Paris, Ser. I}, vol. 346, pp.
  589--592, 2008.

\bibitem{EM08}
Y.~C. Eldar and M.~Mishali, ``Robust recovery of signals from a union of
  subspaces,'' \emph{arXiv.org 0807.4581; submitted to {IEEE} Trans. Inform.
  Theory}, Jul. 2008.

\bibitem{SignsRIP}
R.~Baraniuk, M.~Davenport, R.~DeVore, and M.~Wakin, ``A simple proof of the
  restricted isometry property for random matrices,'' \emph{Const. Approx.},
  2007.

\bibitem{BD08}
T.~Blumensath and M.~E. Davies, ``Sampling theorems for signals from the union
  of linear subspaces,'' \emph{preprint}, 2007.

\end{thebibliography}

\end{document}